# Temperature Evolution of Quasi-one-dimensional $C_{60}$ Nanostructures on Rippled Graphene

Chuanhui Chen, Husong Zheng, Adam Mills, James R. Heflin, and Chenggang Tao[*]

*Department of Physics, Virginia Tech, Blacksburg, Virginia 24061, United States*

[*] Correspondence should be addressed to Chenggang Tao (E-mail: cgtao@vt.edu).

## Abstract

We report the preparation of novel one-dimensional (1D) $C_{60}$ nanostructures on rippled graphene. Through careful control of the subtle balance between the linear periodic potential of rippled graphene and the $C_{60}$ surface mobility, we demonstrate that $C_{60}$ molecules can be arranged into a 1D $C_{60}$ chain structure with widths of two to three molecules. At a higher annealing temperature, the 1D chain structure transitions to a more compact hexagonal close packed quasi-1D stripe structure. This first experimental realization of 1D $C_{60}$ structures on graphene may pave a way for fabricating new $C_{60}$/graphene hybrid structures for future applications in electronics, spintronics and quantum information.

Graphene, a single atom thick layer of sp$^2$ hybridized carbon atoms, is a unique two-dimensional (2D) material that exhibits fascinating physical properties and has numerous potential applications.[1-3] To understand the physical and chemical properties of graphene and explore its potential applications, significant research efforts have recently been devoted to investigate the adsorption and desorption of various molecules on graphene, and to further optimize and control the molecule/graphene hybrid structures. For these studies, graphene grown by chemical vapor deposition (CVD) on metals is emerging as an ideal platform, which is also one of the most promising candidates for practical applications of graphene that require scalable production of graphene. [4,5]

Buckminsterfullerene, $C_{60}$, is a closed-cage structure composed of 60 carbon atoms linked by single and double bonds to form a hollow sphere. Due to its many exciting chemical and physical properties, numerous applications have been demonstrated in single-electron transistors, superconductivity, and photovoltaics. [6,7] Due to its single element composition and highly symmetrical shape, $C_{60}$ often serves as a model system for understanding molecule-surface interactions. $C_{60}$ on epitaxial graphene, a system exclusively formed of elemental carbon, provides an ideal model system to study interactions at the molecular level, electromigration, molecular electronics and spintronics. On most metal or graphite surfaces, $C_{60}$ molecules self-assemble into isotropic thermodynamically favorable hexagonal close packed monolayers. [8-11]

A particularly interesting arrangement of $C_{60}$ molecules is quasi-one-dimensional (1D) structures. Highly ordered 1D molecular configurations are excellent model systems and prototypes of 1D quantum confinement of electronic states, and thus have potential importance in electronic nanodevices, spintronics and solid-state quantum computation. [12-15] However, quasi-1D $C_{60}$ nanostructures have been rarely realized experimentally due to their highly anisotropic configuration. A few previously reported experimental results include $C_{60}$ peapods by embedding $C_{60}$ molecules in carbon nanotubes, 1D $C_{60}$ structures aligned along step edges on vicinal surfaces of metal single crystals, and $C_{60}$ chains on self-assembled molecular layers.[16-19] For CVD graphene, periodically linear rippled structures have been previously observed on various growth substrates.[20,21] Here we experimentally realize quasi-1D $C_{60}$ nanostructures on rippled graphene by utilizing the linear periodic potential in graphene as a template. Through carefully controlling the subtle balance between the weak periodic potential and the surface mobility of $C_{60}$, $C_{60}$ molecules on rippled graphene are arranged into a novel 1D $C_{60}$ chain structure with widths ranging from two to three molecules wide. At a higher annealing temperature, the 1D chain structure transitions to a more compact hexagonal close packed quasi-1D stripe structure. To the best of our knowledge, this is the first experimental realization of 1D $C_{60}$ structures on graphene. Our experimental results may pave a way for fabricating new $C_{60}$/graphene hybrid structures for future applications in electronics, spintronics (e.g. endohedral fullerenes with magnetic spins can be aligned in 1D configurations), and quantum information.

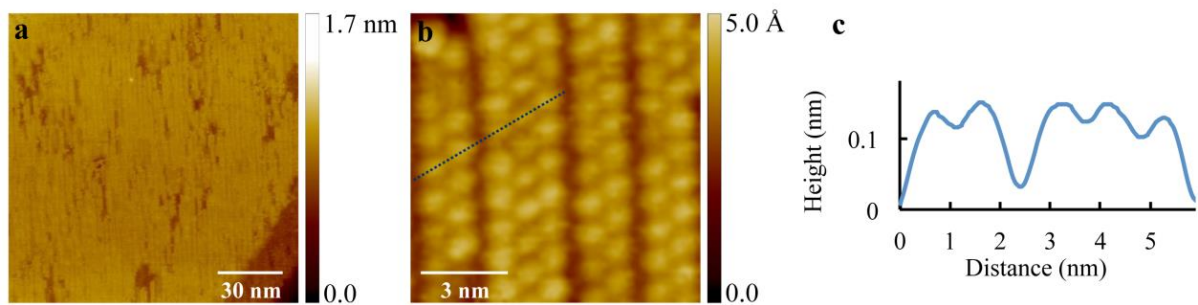

**Fig. 1. 1D bimolecular and trimolecular $C_{60}$ chains after annealing at 423 K.** (a) Large scale STM image of $C_{60}$ on graphene showing well ordered 1D structure on Cu facets ($V_s$ = 1.90 V, I = 0.40 nA). (b) High-resolution image of bimolecular and trimolecular $C_{60}$ chains. Within the chains, the $C_{60}$-$C_{60}$ intermolecular spacing is ~ 1.0 nm, and the interchain distance, defined as the distance between the centers of adjacent $C_{60}$ rows belonging to neighboring chains, is 1.23 nm ($V_s$ = 1.95 V, I = 0.50 nA). (c) A line profile along the close packed orientation as marked with the dashed blue line in (b).

$C_{60}$ molecules were deposited onto graphene grown by the CVD method on Cu foil at the

background pressure below $2.0 \times 10^{-8}$ mbar (more details are in the Method section). After post-deposition annealing at 423 K, we observed small atomically flat facets with typical sizes ranging from 50 nm to 200 nm on the surface. These facets can be classified into two predominant types based on the distinguishable surface characteristics, labeled as type A and type B (see the Supporting Information Figs. S1a and S1b). On type A facets, $C_{60}$ molecules form hexagonal close packed monolayers and islands (see the Supporting Information Fig. S3a). On type B facets we observed a characteristic quasi-1D feature of $C_{60}$ molecules as shown in the large-area overview STM image in Fig. 1a. A closer inspection of the high-resolution STM images reveals the details of this 1D structure featured in Fig. 1b, in which each bright protrusion represents a $C_{60}$ molecule. Typically, the 1D chains consist of two or three $C_{60}$ rows, referred to as bimolecular and trimolecular $C_{60}$ chains, respectively. Within a chain, the $C_{60}$-$C_{60}$ intermolecular spacing is $1.00 \pm 0.01$ nm, indicating that $C_{60}$ molecules arrange in a close packed manner. From the line profile shown in Fig. 1c that corresponds to the dashed line marked in Fig. 1b, the isolation between adjacent $C_{60}$ chains can be clearly seen from the gap between the second and the third peaks. Here, the second and the third peaks represent the $C_{60}$ molecules belonging to neighboring chains. The measured average interchain distance, defined as the distance between the centers of adjacent $C_{60}$ rows belonging to neighboring chains, is $1.23 \pm 0.02$ nm, as schematically shown Fig. 3a. In our measurements, we observed that the chains consist exclusively of two or three $C_{60}$ rows. From statistical analysis, the bimolecular chains are shown to be more numerous, with a ratio of number of bimolecular to trimolecular chains of about 2.0. As observed in large-scale STM images, the chains are usually arranged in a 3-2-2 or 2-3-2 manner. In our measurements, we observed that the orientation of these $C_{60}$ chains was exclusively along the same direction, nearly parallel to the bending axis. In the chain structure we also observed some junctions where a bimolecular segment transitions into a trimolecular segment, or vice versa.

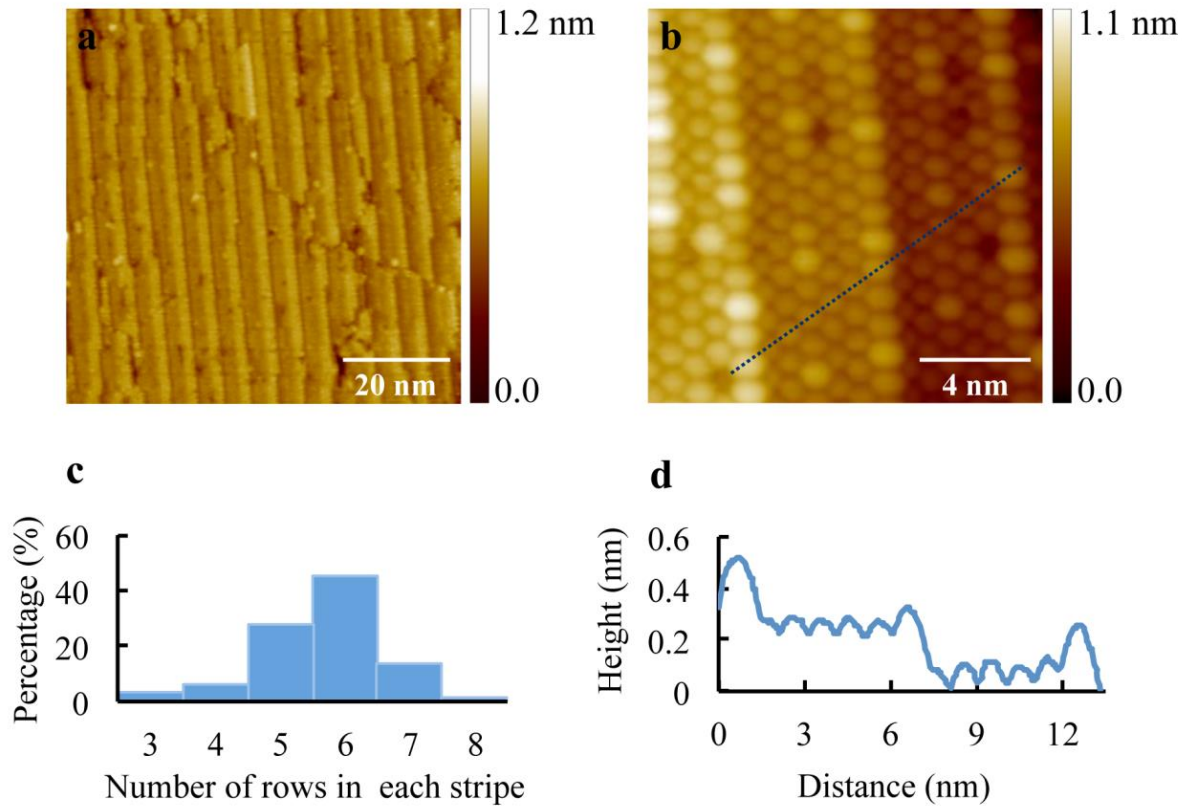

**Fig. 2. Self-assembled quasi-hexagonal close packed 1D $C_{60}$ stripe structure on rippled graphene after annealing at 487 K.** (a) Large-scale STM image of $C_{60}$ on graphene showing quasi-hexagonal close packed 1D stripe structure ($V_s$ = 2.20 V, I = 0.20 nA). (b) High-resolution image of the quasi-hexagonal close packed 1D stripes ($V_s$ = 2.40 V, I = 0.20 nA). (c) Normalized distribution of $C_{60}$ stripe widths in terms of the number of rows. (d) A line profile along the close packed orientation as marked with the dashed blue line in (b).

In order to investigate the temperature evolution of the $C_{60}$ nanostructures on graphene, we performed subsequent annealing at a higher temperature, 487 K, for 2 hours. After annealing at this temperature, the $C_{60}$ chain structure disappeared and a new more compact hexagonal close packed quasi-1D stripe structure emerged, as shown in Fig. 2a. The zoom in topographic STM image in Fig. 2b reveals hexagonal close packed quasi-1D stripes with widths of three and six $C_{60}$ rows and three $C_{60}$ rows. A noticeable difference from the $C_{60}$ chain structure is that these quasi-1D stripes are formed on staggered narrow terraces, which are between nearly straight and parallel step edges (Figs. 2b and 2d). In the stripe structure there is no spacing between neighboring stripes. In contrast, the quasi-1D stripe structure appears more crowded than the close packed structure. The lateral inter-row distance between the two rows adjacent to the step edges, on the upper and lower terraces respectively, is $0.75 \pm 0.01$ nm, as schematically shown in the ball models in Figs. 3b and 3d. This high-density arrangement seems to accommodate the underlying narrow terraces that formed after annealing at 487 K. On the narrow terraces, $C_{60}$ molecules are arranged in a close packed manner with a characteristic $C_{60}$-$C_{60}$ intermolecular spacing of $1.00 \pm 0.02$ nm. We also noticed that the $C_{60}$ row near the step edge on the upper terrace is with a measured height higher than the other rows on the same terrace as shown in Figs. 2b and 2d. Different from the dominance of $C_{60}$ chains with widths of two or three $C_{60}$ rows, the width of the stripes varies from three to eight rows of $C_{60}$. The most common stripes have a width of 6 $C_{60}$ rows. Fig. 2c is the normalized distribution of the $C_{60}$ stripe width in terms of the number of $C_{60}$ rows. The 6-row stripes are the most common stripe structure, with a probability of 45%, and 5-row stripes are the second most likely stripe structure. The 8-row and 3-row stripes are minority groups. The lateral periodicity of these typical 6-row stripes is $5.08 \pm 0.02$ nm, almost exactly the same as the lateral spacing of a bimolecular chain plus a neighboring trimolecular chain in the chain structure, which is also $5.08 \pm 0.02$ nm. The stripes are nearly parallel to each other and the orientation of the stripes is along the same orientation with the $C_{60}$ chains. Similar to the chain structure, there are junction regions in these quasi-1D stripes. Another detail worth noticing is that the down hill direction of the stripes is determined by the boundaries of the facets. The protrusion boundaries are the downhill side while the depression boundaries correspond to the uphill side (see Figs. S2a and S2b in the Supporting Information).

To compare the chain structure and the stripe structure more clearly and visually, we try to use a 3D model to explain it. Figs. 3a and 3c is the schematic model for the bimolecular and trimolecular $C_{60}$ chains with dark blue spheres representing $C_{60}$ molecules and green spheres representing carbon atoms of underlying graphene. The single $C_{60}$ row on the left side is the adjacent row in the next bimolecular or trimolecular chain. The width of a bimolecular cell (chain plus the interchain spacing) plus a trimolecular cell is 5.08 ± 0.02 nm. From these models, we can clearly see the large spacing (1.23 nm) between adjacent chains (Fig. 3a). In the stripe structures, the inter-row spacing between the $C_{60}$ rows adjacent to step edges from both sides is 0.75 nm, narrower than the other $C_{60}$ rows arrange in a hexagonal close packed manner (Figs. 3b and 3d).

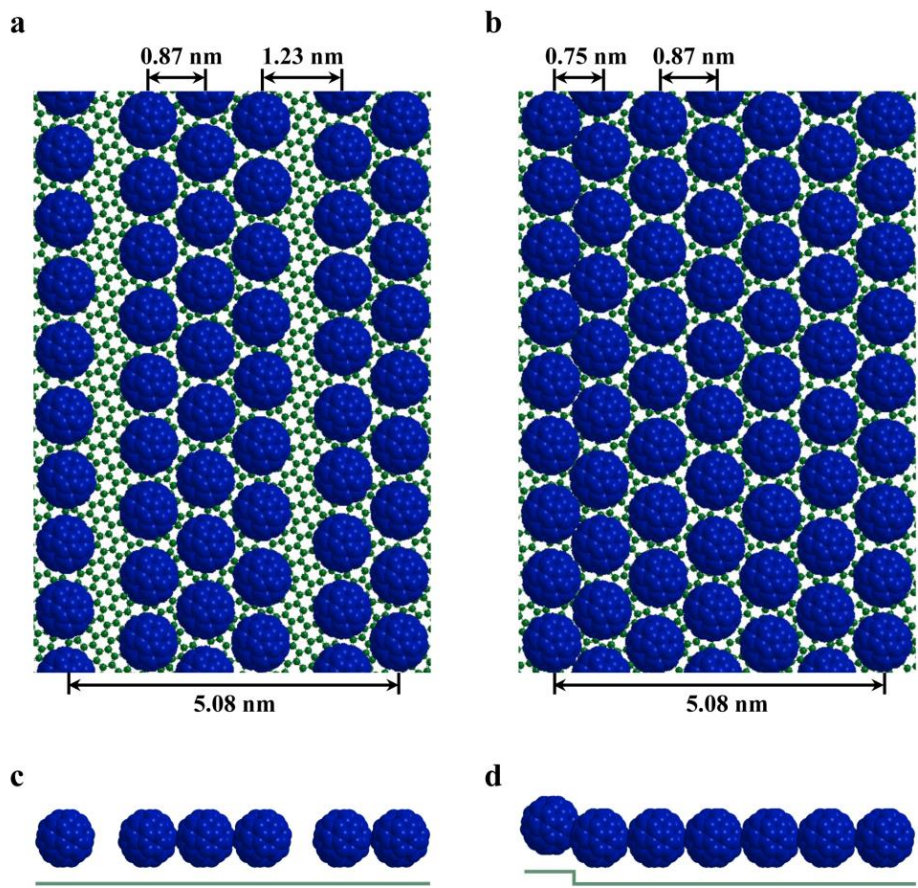

**Fig. 3. Schematic models of the $C_{60}$ chain and stripe structures showing the graphene layer (small dark green balls) and the $C_{60}$ layer (dark blue space filling spheres) on top of the graphene layer.** (a) and (c) Bimolecular and trimolecular $C_{60}$ chains on graphene after annealing at 423 K. (b) and (d) $C_{60}$ stripe structure with 6-row width after annealing at 487 K.

To gain detailed insight into the growth mechanisms of the quasi-1D $C_{60}$ chain and stripe structures, we used STM to characterize the underlying graphene of the empty areas adjacent to the $C_{60}$ quasi-1D structures, and the bare graphene surface before $C_{60}$ deposition and after $C_{60}$ desorption. After annealing the temperature up to 578 K, all $C_{60}$ molecules were desorbed from the surface, and the surface was restored to bare graphene on Cu. The desorption temperature is close to the desorption temperature of $C_{60}$ molecules on highly ordered pyrolytic graphite (HOPG). [22] The similar deposition temperature indicates a similar activation energy, which is 1.69 eV for $C_{60}$ on graphite based on the thermal desorption spectroscopy study. [22] Atomic-resolution images taken on small flat facets of the substrates, where the quasi-1D $C_{60}$ chains and stripes are grown, show a well-defined linear periodic graphene rippled superstructure with an orientation close to the chain and stripe orientation. Fig. 4a shows the typical linear periodic modulation of graphene with the periodicity of 0.75 ± 0.01 nm and with the same orientation of the observed $C_{60}$ chains and stripes, as indicated by the blue arrow in Fig. 4b. The linear periodic structure as shown in Figs. 4a and 4b is attributed to the case where the underlying Cu facet is Cu(100). To confirm that, we checked two characteristic surface features. The step edges have a height of ~ 0.18 nm, which is the characteristic step height of monoatomic steps on Cu (100) surface. [23,24] This kind of linear modulation of graphene due to the square lattice symmetry of the Cu(100) surface has been experimentally observed before. [20,21,23] The superposition of a graphene lattice on a Cu(100) surface with various angles produces linear rippled patterns with different periodicity. [20,21,23] At some certain angle, there will be maze-like reconstruction of graphene on Cu(100). [25] However, superstructures of graphene on Cu(111) surface have characteristic six-fold symmetry, [26-31] which can be distinguished easily from the linear periodic ripple structure in our experiments. In a few cases, linear superstructures of graphene on Cu(111) were also observed, but they usually have a much larger periodicity and disappear after annealing to 500 K. [32] The graphene ripples in our experiments survived after annealing to 693 K. From all these observations, it is reasonable to conclude that the Cu surface underneath rippled graphene is Cu(100). Based on the periodicity of the ripples, we are able to calculate the relative angle between graphene and the underlying Cu(100) surface. Fig. 4c is a suggested atomic model for the rippled graphene. The zigzag direction of graphene has an angle of 13.0° relative to the $[01\bar{1}]$ direction of the underlying Cu(100). The calculated periodicity is 0.80 nm and the orientation of ripples is with a relative angle of 15.2° to the zigzag direction of graphene, consistent with the measured angle of 15.0° ± 0.5°. The calculated periodicity is slightly larger than the measured periodicity of 0.75 nm. This difference may attribute to the out of plane distortion of graphene ripples, as shown in the line profile in Fig. 4d.

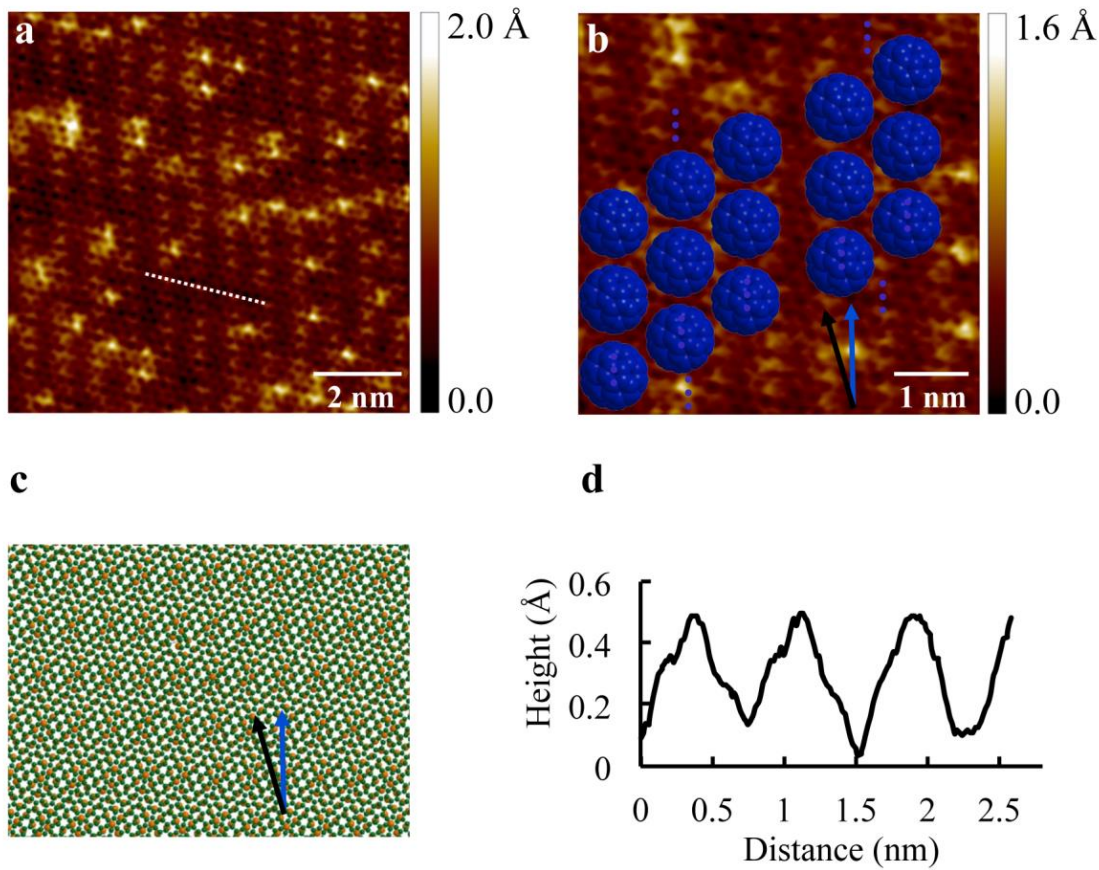

**Fig. 4. STM topography images of rippled graphene on Cu.** (a) Large area STM topographic image of the rippled graphene showing well-defined linear periodic modulation with a 0.75 nm spatial modulation frequency ($V_s$ = 0.80 V, I = 1.0 nA). (b) Magnified STM image of the rippled graphene with the ball model of $C_{60}$ molecules in the arrangement of a bimolecular (right) and trimolecular (left) chain with the orientation along the ripples. The blue arrow indicates the orientation of the linear graphene ripples and the $C_{60}$ chains, with a measured angle of 15.0° ± 0.5° relative to the zigzag direction of graphene, indicated by the black arrow. The periodicity of the linear modulation is 0.75 ± 0.01 nm ($V_s$ = 0.80 V, I = 1.0 nA). (c) Suggested ball and stick model of the overlay of graphene on Cu(100). The zigzag direction of graphene is rotated by 13.0° with respect to the [01$\bar{1}$] direction of the Cu(100) underneath. The green balls represent carbon atoms of graphene, and the yellow balls represent Cu atoms of the underlying Cu surface. The blue arrow indicates the orientation of the linear moiré pattern, and the black arrow indicates the zigzag direction of graphene. (d) Line profile along the zigzag orientation of graphene (marked as a dashed white line in (a)) showing the periodic modulation.

The orientation of the graphene ripples is fully consistent with that of the $C_{60}$ chain structure. Based on the rippled graphene structure, it is possible for us to explain the formation of the chain structure from the point of view of mechanical stability. [33] The most common morphology for the $C_{60}$ on flat graphene is the hexagonal close-packed structure due to the weak diffusion barrier on flat graphene and the much stronger $C_{60}$-$C_{60}$ interaction. [34,35] On rippled graphene, the linear periodic potential provides a template for $C_{60}$ to arrange into more stable 1D configurations. The linear periodicity of rippled graphene, 0.75 ± 0.01 nm, is smaller than the typical inter-row distance of 0.87 nm in the $C_{60}$ closed packed arrangement. $C_{60}$ molecules randomly occupy the valley sites of the linear periodic potential at the initial stage of growth. It is possible for $C_{60}$ molecules to settle down in the adjacent ripples at a less stable state because $C_{60}$ molecules are forced to stay slightly off the valley sites. In this way, two or three adjacent linear graphene ripples are filled up. This is consistent with the observation that almost all $C_{60}$ chains consist of two or three rows. Based on a geometrical and energetic argument, it is easy to see that chains with 4 or more rows are less likely to form. Compared with the formation of the bimolecular chains, the possibility to form trimolecular chains (both sides need a $C_{60}$ molecule sitting down around the middle $C_{60}$ molecule) is about half the probability of the formation of the bimolecular chains, which only need one side to be occupied with $C_{60}$ molecules. From a Monte Carlo statistical approach, $C_{60}$ molecules adsorbed on such a linear graphene sheet will be stabilized in one of the valleys [33], and if a second or a third molecular row packs closely to the stabilized molecular row, the chain structure consequently forms. We expect that $C_{60}$ molecules with less thermal energy can form monomolecular chains. That is one of the possible directions to further tune and control the formation of these 1D structures.

In order to understand how the bimolecular and trimolecular chains transform into the $C_{60}$ stripe structure after annealing at 487 K, we will start by considering the diffusion of Cu atoms underneath graphene and the arrangement of $C_{60}$ molecules on graphene. When annealing at a higher temperature of 487 K, $C_{60}$ molecules have a higher probability to overcome the potential energy difference between the hilltop and the valley sites of rippled graphene and rearrange into chains with larger width due to the higher thermal excitation. The junction structure in the bimolecular and trimolecular chains probably provides diffusion channels for $C_{60}$ molecules to transfer between adjacent chains resulting in emergent chains with larger widths. Considering that the arrangement of bimolecular and trimolecular chains is usually 2-3-2 or 3-2-2 as described above, it is reasonable to assume that the emergent chains with width of five $C_{60}$ rows are the configuration with highest probability to form, compared with chains of other widths. The distribution of chains widths and stripe widths may imply that at the annealing temperature of 487 K, the thermal excitation can only drive $C_{60}$ molecules to cross over one or two energy barriers, but not more. When adjacent bimolecular and trimolecular chains merge into a 5-row chain, the interchain spacing between the merged 5-row chain to its neighboring chains can be generated from the initial interchain spacing in the chain configuration (1.23 nm), which is 1.59 nm. This spacing is slightly smaller than the interchain spacing of 1.73 nm required for a new $C_{60}$ row to fill in the empty space. It has been previously reported that Cu atoms can easily diffuse underneath a graphene sheet even at room temperature. [23] These detailed observations suggest that copper atoms can diffuse along smooth step edges and kinked sites, and even dissociate from step edges, indicating a weak interaction between the graphene sheet and the underlying copper atoms. [23] Considering the boundaries of the facets (see the Supporting Information Fig. S2a), it is likely that the depression boundaries provide a source of copper atoms while the protrusion boundaries provide a drain. Here the surface reconstruction due to diffusion of underlying copper atoms provides the possibility for another $C_{60}$ row to form along the straight step edges and results in quasi-hexagonal close packed 1D $C_{60}$ stripes with the typical width of six rows. The origin of $C_{60}$ stripes with other number of rows can be attributed to the emergence of different combinations of $C_{60}$ chains, such as two bimolecular chains emerging into a stripe with five $C_{60}$ rows.

In both $C_{60}$ chain and stripe structures (Figs. 1 and 2), we notice that a few $C_{60}$ molecules appear higher than others, and the relative height shows bias dependence. The difference is possibly due to the defect sites in the underlying graphene that appear as bright sites in the STM images of graphene (Fig. 4a).

In summary, our results demonstrate that novel quasi-1D $C_{60}$ structures can be realized on rippled graphene. These quasi-1D structures are controllable by tuning the temperature, from a chain structure with widths of two or three $C_{60}$ molecules, and then to a quasi-hexagonal close packed 1D stripe structure. In our experiments, temperature is the main tunable parameter. Based on the subtle balance between the $C_{60}$-$C_{60}$ intermolecular and the $C_{60}$-graphene interactions, as well as the size of the graphene ripples due to the moiré patterns, we expect that the quasi-1D $C_{60}$ structures on rippled graphene can be further tuned and controlled by varying the size of the ripples, the coverage of adsorbed $C_{60}$ molecules, and the temperature.

## Methods

All experiments were carried out in an ultra-high vacuum (UHV) scanning tunneling microscope system (Omicron LT-STM). Before $C_{60}$ deposition, the graphene grown by the CVD method on Cu foil was annealed for 12 hours at 673 K in a preparation chamber with a base pressure of low $10^{-10}$ mbar. $C_{60}$ powder (MER Corporation, 99.5% purity) was loaded into the homemade Knudsen cell mounted on the load lock of the LT-STM system. The $C_{60}$ source was degassed at a pressure of low $10^{-8}$ mbar before deposition. $C_{60}$ molecules were then sublimated at a deposition rate of ~ 0.9 ML/min with the background pressure below $2.0 \times 10^{-8}$ mbar. During the deposition process, the substrate was kept at room temperature. The samples were then subsequently annealed at 423 K, 487 K, and 578 K respectively in the preparation chamber of the STM system with the base pressure at low $10^{-10}$ mbar. All the STM measurements were performed at room temperature in the STM characterization chamber with a base pressure of about $3.0 \times 10^{-11}$ mbar.

**Competing financial interests:**

The authors declare that they have no competing financial interests.